\documentclass[fleqn,usenatbib]{mnras}
\usepackage{graphicx}
\usepackage{amsmath}
\usepackage{newtxtext,newtxmath}

\usepackage[utf8]{inputenc} 
\usepackage{kotex} 
\usepackage{hyperref}

\usepackage{bm} 
\usepackage{natbib} 

\usepackage{booktabs}
\usepackage{float}
\usepackage{caption}    
\usepackage{subcaption} 
\usepackage{enumitem}   
\usepackage{ragged2e}   
\usepackage{tabularx}   

\newcommand{\oo}{\omega_{0}}
\newcommand{\oa}{\omega_{a}}

\title[Impact of $\Omega_{m0}$ Prior on DE Parameters]{The Impact of $\Omega_{m0}$ Prior Bias on Cosmological Parameter Estimation: Reconciling DESI DR2 BAO and Pantheon+ SNe Data}

\author[Seokcheon Lee]{Seokcheon Lee \thanks{E-mail: skylee@skku.edu} \\
Department of Physics, Institute of Basic Science, Sungkyunkwan University, Suwon 16419, Korea}

\date{Accepted XXX. Received YYY; in original form ZZZ}
\pubyear{2025}

\begin{document}

\maketitle

\begin{abstract}
Recent cosmological parameter analyses combining DESI DR2 Baryon Acoustic Oscillation (BAO) data with external probes, such as Pantheon+ Supernovae (SNe) observations, have reported deviations of the dark energy equation-of-state parameters ($\omega_0, \omega_a$) from the standard $\Lambda$CDM model predictions ($\omega_0=-1, \omega_a=0$). A notable aspect of these results is the role of $\Omega_{m0}$ prior information from SNe, which is known to exhibit tension with BAO-only constraints. In this study, we rigorously investigate this effect through a statistical analysis using 1000 mock DESI DR2 BAO data realizations. We demonstrate that the strong degeneracy between $\omega_0$, $\omega_a$, and $\Omega_{m0}$ causes significant biases in the estimated dark energy parameters when the $\Omega_{m0}$ prior mean deviates from its true underlying value. Specifically, applying an $\Omega_{m0}$ prior mean of 0.33 (consistent with some SNe-only constraints) to mock data, assuming a true $\Lambda$CDM universe ($\Omega_{m0}=0.30, \omega_0=-1, \omega_a=0$), yields biased estimates such as $\omega_0 \approx -0.82 \pm 0.06$ and $\omega_a \approx -0.82 \pm 0.4$. This systematic shift, driven by the $\Omega_{m0}$ prior, moves the estimated parameters towards the non-$\Lambda$CDM region, offering a qualitative resemblance to outcomes reported in current combined DESI DR2 BAO + Pantheon+ SNe analyses (e.g., $\omega_0 = -0.888^{+0.055}_{-0.064}$, $\omega_a = -0.17 \pm 0.46$). Our findings suggest that these observed non-$\Lambda$CDM parameters may largely arise from statistical biases due to $\Omega_{m0}$ prior tensions between datasets. This study proposes a potential resolution to current cosmological tensions without necessarily invoking new physics.
\end{abstract}

\begin{keywords}
cosmology: observations -- dark energy -- large-scale structure of Universe -- methods: statistical
\end{keywords}

\section{Introduction}
\label{sec:Int}

The discovery of the accelerating expansion of the Universe has fundamentally transformed modern cosmology. This phenomenon is most commonly attributed to the presence of dark energy (DE)~\cite{SupernovaCosmologyProject:1998vns,SupernovaSearchTeam:1998fmf}, though alternative explanations involving modifications to general relativity or other new physics are actively explored~\cite{Clifton:2011jh,Ishak:2018his,Perivolaropoulos:2021jda,CosmoVerseNetwork:2025alb}. Unveiling the nature of DE—often characterized by its equation of state (EOS) parameters $\omega_0$ and $\omega_a$—remains one of the central challenges in contemporary cosmology~\cite{Cooray:1999da,Astier:2000as,Chevallier:2000qy,Efstathiou:1999tm,Jassal:2005qc,Barboza:2008rh,Lee:2011ec}. Precise constraints on these parameters are essential for distinguishing the standard cosmological constant model ($\Lambda$CDM, where $\omega_0 = -1$ and $\omega_a = 0$) from more exotic dynamical dark energy scenarios. Complementary analyses using multiple parameterizations of $\omega(z)$ or non-parametric reconstructions have been explored in recent works~\cite{Giare:2024oil,DESI:2025fii},  offering an additional perspective on the evolving dark energy constraints from Dark Energy Spectroscopic Instrument (DESI) and other probes.

Large-scale structure surveys such as the DESI have become crucial in this effort by measuring the Baryon Acoustic Oscillation (BAO) feature across cosmic time with high precision~\cite{DESI:2024mwx,DESI:2025fxa,DESI:2025zgx}. The recently released DESI DR2 BAO dataset offers some of the tightest constraints to date. To further improve parameter estimation, it is common to combine BAO data with complementary cosmological probes, such as Type Ia supernovae (SNe) from the Pantheon+ sample, in order to break degeneracies and increase overall sensitivity.

However, recent joint analyses combining DESI DR2 BAO and Pantheon+ SNe data have revealed a consistent deviation of the DE EOS parameters $\omega_0$ and $\omega_a$ from their fiducial $\Lambda$CDM values~\cite{DESI:2025zgx}. This tension has prompted growing interest in whether new physics may be required. At the same time, a notable discrepancy has emerged in the matter density parameter, $\Omega_{m0}$, inferred from each dataset: DESI DR2 BAO favors $\Omega_{m0} \approx 0.2975 \pm 0.0086$~\cite{DESI:2025zgx}, whereas Pantheon+ SNe indicates a higher value, $\Omega_{m0} \approx 0.334$~\cite{Brout:2022vxf}. This tension in $\Omega_{m0}$ is critical, as it propagates into derived constraints on the DE EOS when the datasets are combined~\cite{Pedrotti:2024kpn,Wang:2025vfb,Lee:2025hjw,Patel:2024odo,Colgain:2024mtg,Park:2025azv,Sakr:2025daj,Colgain:2025nzf,Gonzalez-Fuentes:2025lei}.

In this work, we systematically examine the hypothesis that such $\Omega_{m0}$ tension can induce a statistical bias in the inferred values of $\omega_0$ and $\omega_a$~\cite{Patel:2024odo}. We generate 1000 mock realizations of DESI DR2 BAO data under the assumption of a fiducial $\Lambda$CDM cosmology ($\Omega_{m0} = 0.3$, $\omega_0 = -1.0$, $\omega_a = 0.0$) and assess how varying the mean of the $\Omega_{m0}$ prior influences the resulting DE parameter constraints. This approach allows us to disentangle whether current deviations from $\Lambda$CDM reported in the literature are indicative of true physical phenomena or are instead artifacts of mismatched priors between datasets. Our results offer a compelling case for the latter, and thus provide a possible resolution to recent observational tensions without invoking new physics.

\section{Methodology}
\label{sec:Meth}

Our analysis followed a rigorous procedure, comprising the construction of a fiducial cosmological model, evaluation of BAO observable sensitivities, simulation of mock datasets with covariance structure, and parameter estimation via Bayesian inference. The detailed steps are outlined below.

\subsection{Fiducial Model and BAO Sensitivities}
\label{subsec:sensitivities}

We adopt a fiducial $\Lambda$CDM cosmology with $\Omega_{m0} = 0.3$, $\omega_0 = -1.0$, and $\omega_a = 0.0$ as the baseline for our analysis. This serves as the underlying truth model for mock data generation and for computing parameter sensitivities.

The constraining power of BAO observables depends on their sensitivity to cosmological parameters and on the observational uncertainties, encapsulated in the covariance matrix. To evaluate sensitivities, we use the standard FLRW metric in a Universe with evolving dark energy EOS $\omega(z) = \omega_0 + \omega_a \frac{z}{1+z}$. The relevant theoretical expressions are:
\begin{align}
&E(z) \equiv \frac{H(z)}{H_0} = \nonumber \\
&\sqrt{\Omega_{r0} (1+z)^4 + \Omega_{m0} (1+z)^3 + (1 - \Omega_{r0} - \Omega_{m0})  f_{\textrm{DE}}(z) }, \label{eq:E_z} \\
&\textrm{where} \quad f_{\textrm{DE}}(z) = (1+z)^{3(1+\omega_0+\omega_a)}  e^{-3 \omega_a \frac{z}{1+z}} \,, \nonumber \\
& r_d = \frac{c}{H_0} \int_{z_d}^{\infty} \frac{dz}{E(z) \sqrt{3(1 + R(z))}}, \quad R(z) = \frac{3 \Omega_{b0}}{4 \Omega_{\gamma 0}} \frac{1}{1+z}, \label{eq:rd_def} \\
& D_H(z) = \frac{c}{H_0} \frac{1}{E(z)}, \quad D_M(z) = \frac{c}{H_0} \int_0^z \frac{dz'}{E(z')}, \nonumber \\ 
& D_V(z) = \left[ z D_M^2(z) D_H(z) \right]^{1/3}, \label{eq:distances}
\end{align}
with the derived combinations:
\begin{align}
& \frac{D_M(z)}{D_H(z)} = E(z) \int_0^z \frac{dz'}{E(z')}, \quad
\frac{D_H}{r_d} = \left[ E(z) \int_{z_d}^{\infty} \frac{dz}{E(z) \sqrt{3(1 + R(z))}} \right]^{-1}, \nonumber \\
& \frac{D_M}{r_d} = \frac{\int_0^z dz'/E(z')}{\int_{z_d}^{\infty} dz/[E(z) \sqrt{3(1 + R(z))}]}, \nonumber \\
& \frac{D_V}{r_d} = \frac{z^{1/3} \left( \int_0^z \frac{dz'}{E(z')} \right)^{2/3}}{E^{1/3}(z) \int_{z_d}^{\infty} dz/[E(z) \sqrt{3(1 + R(z))}]}
\label{eq:observables} \,,
\end{align}
where the symbols and quantities appearing in Eq.~\eqref{eq:observables} are defined as follows:
the density contrasts $\Omega_{r0}$ (radiation), $\Omega_{m0}$ (matter), $\Omega_{b0}$ (baryons), $\Omega_{\gamma0}$ (photons),  $r_d$ (sound horizon at the drag epoch $z_d$), $D_H\!=\!c/H(z)$, $D_M$ (comoving angular-diameter distance), 
and $D_V$ (volume-averaged BAO distance). All parameter values used for sensitivities and mocks are stated next to the corresponding equations.

We calculate the absolute sensitivities $|\partial O / \partial \theta|$ of the BAO observables $\mathbf{O} = (D_V/r_d, D_M/D_H, D_M/r_d, D_H/r_d)$ with respect to the cosmological parameters $\theta = (\Omega_{m0}, \omega_0, \omega_a)$. These enter the Fisher Information Matrix (FIM), defined as:
\begin{equation}
F_{ij} = \left( \frac{\partial \mathbf{O}}{\partial \theta_i} \right)^T \mathbf{C}^{-1} \left( \frac{\partial \mathbf{O}}{\partial \theta_j} \right),
\end{equation}
where $\mathbf{C}$ is the covariance matrix of the observables.

\begin{figure}
    \centering
    \begin{subfigure}[b]{0.45\textwidth}
        \includegraphics[width=\textwidth]{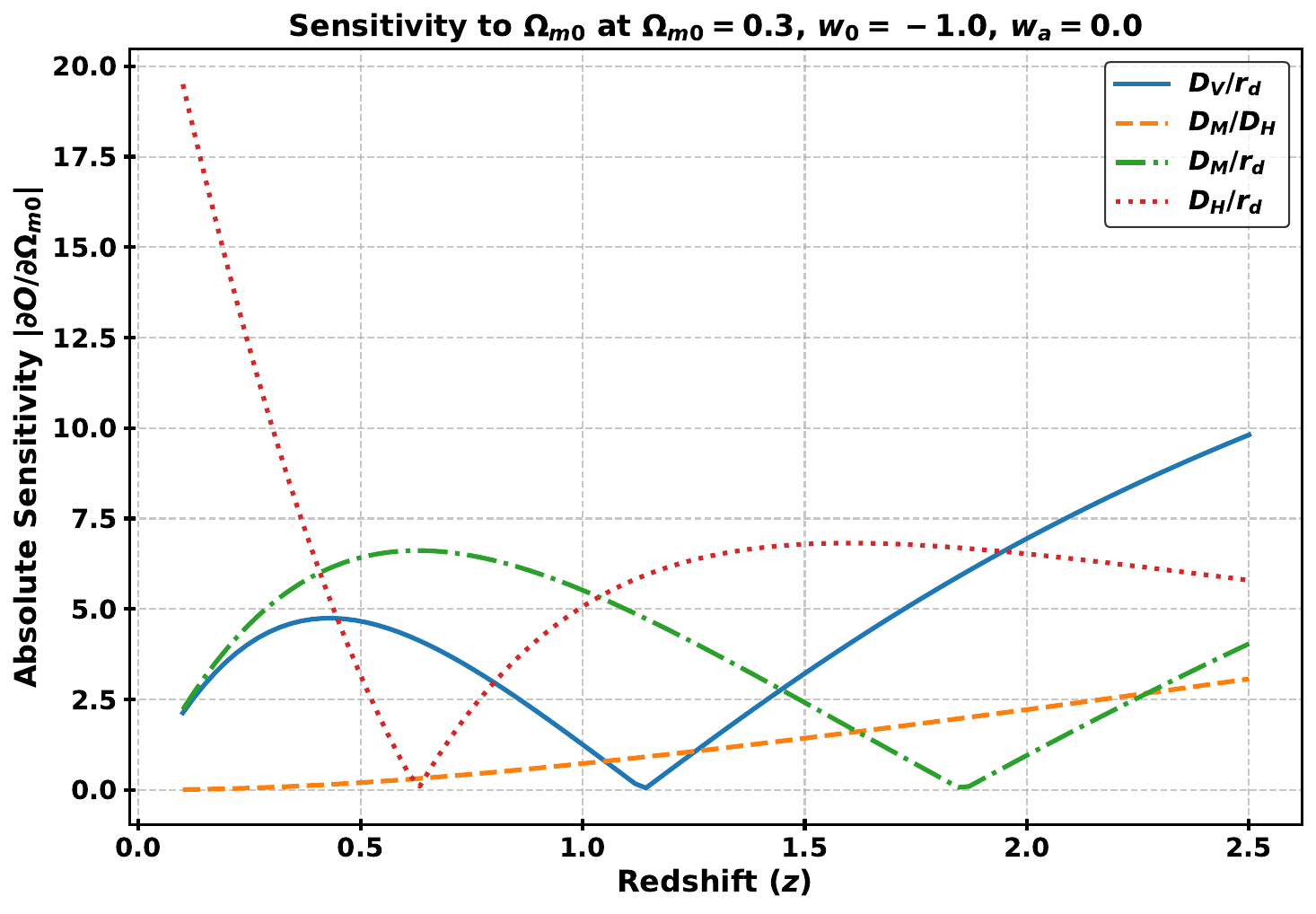}
        \caption{Sensitivity to $\Omega_{m0}$}
    \end{subfigure}
    \hfill
    \begin{subfigure}[b]{0.45\textwidth}
        \includegraphics[width=\textwidth]{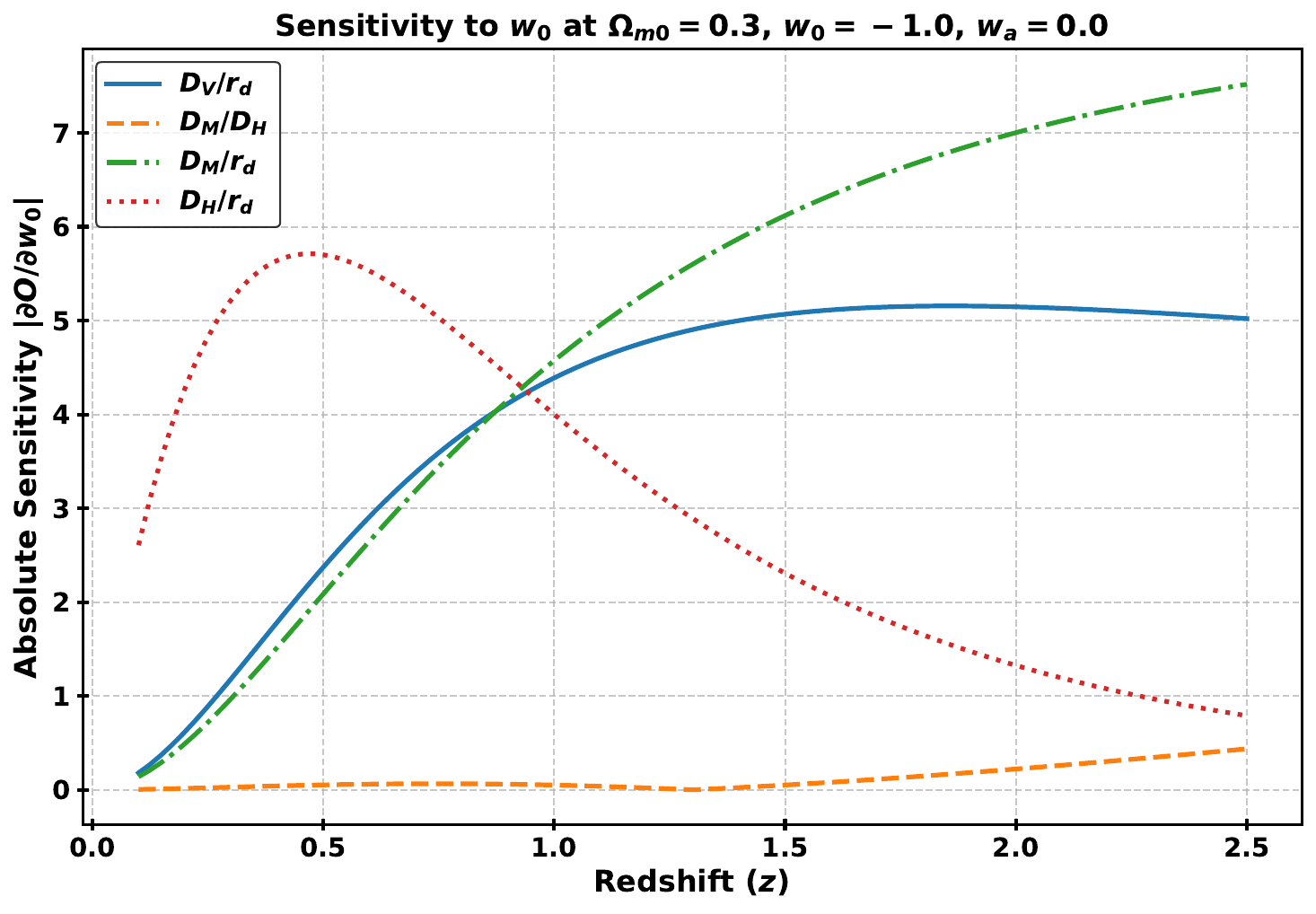}
        \caption{Sensitivity to $\omega_0$}
    \end{subfigure}

    \vspace{1em}

    \begin{subfigure}[b]{0.45\textwidth}
        \includegraphics[width=\textwidth]{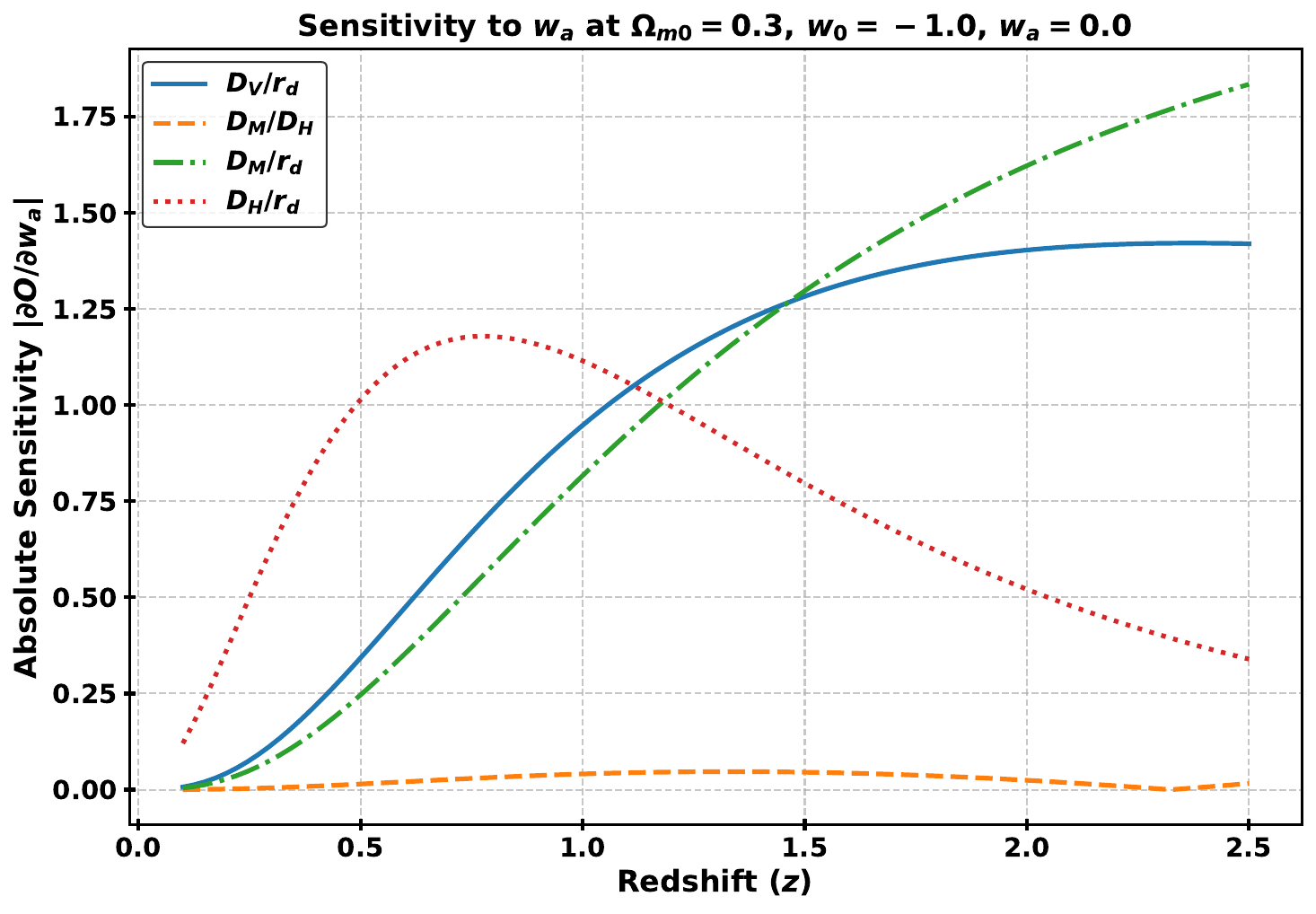}
        \caption{Sensitivity to $\omega_a$}
    \end{subfigure}

    \caption{Absolute sensitivities of BAO observables with respect to $\Omega_{m0}$, $\omega_0$, and $\omega_a$ across redshift. The fiducial cosmology is $\Omega_{m0} = 0.3$, $\omega_0 = -1.0$, $\omega_a = 0.0$.}
    \label{fig:sensitivities}
\end{figure}

Figure~\ref{fig:sensitivities} illustrates the redshift dependence of the sensitivities, quantifying how variations in each parameter impact the BAO measurements.

\subsection{Mock Data and Covariance Matrix Construction}
\label{subsec:mock_data}

We use 1000 DESI DR2 mock realizations as a balance between robustness and computational cost. Although $\Omega_{m0}$ is a free parameter in DESI DR2 BAO analyses, the combination with Cepheid-calibrated Pantheon+ effectively introduces information closely tied to the SH0ES prior on $H_0$. Given the well-known degeneracy between $H_0$ and $\Omega_{m0}$, representing the joint case by an $\Omega_{m0}$ prior is an accurate abstraction of how external distance-ladder information steers the combined posterior. 

To simulate DESI DR2 BAO observations, we construct a block-diagonal covariance matrix $\mathbf{C}$ across redshift bins. Within each bin, observables are ordered as $\mathbf{O} = \left(D_V/r_d, D_M/D_H, D_M/r_d, D_H/r_d\right)$,
with intra-bin covariances defined as:
\begin{align}
&\mathbf{C}_z = \nonumber \\ 
&\begin{pmatrix}
\sigma_{o_1}^2 & \rho_{o_1,o_2} \sigma_{o_1} \sigma_{o_2} & 0 & 0 \\
\rho_{o_1,o_2} \sigma_{o_1} \sigma_{o_2} & \sigma_{o_2}^2 & 0 & 0 \\
0 & 0 & \sigma_{o_3}^2 & \rho_{o_3,o_4} \sigma_{o_3} \sigma_{o_4} \\
0 & 0 & \rho_{o_3,o_4} \sigma_{o_3} \sigma_{o_4} & \sigma_{o_4}^2
\end{pmatrix}.
\label{eq:cov_matrix_block}
\end{align}
Correlations are taken from DESI DR2 Table IV~\cite{DESI:2025zgx}, with:
$\rho_{o_1,o_2} = r_{V,M/H}$, $\rho_{o_3,o_4} = r_{M,H}$. Other intra-bin pairs are assumed uncorrelated, and inter-bin correlations are neglected.

Using this $\mathbf{C}$, we generate 1000 mock BAO datasets from a multivariate Gaussian distribution centered on the fiducial $\Lambda$CDM predictions. These mocks replicate the statistical properties of DESI DR2.

For parameter inference, we perform MCMC analyses combining the BAO mock likelihood with Gaussian priors on $\Omega_{m0}$. The prior mean is scanned from 0.27 to 0.33. For each case, the posterior distributions of $\omega_0$ and $\omega_a$ are obtained. The posterior mean and standard deviation are used as the estimated value and $1\sigma$ uncertainty, respectively, enabling assessment of bias from prior mismatch.

\section{Results and Discussion}
\label{sec:results}

In this section, we present the results of our analysis on the biasing effects of $\Omega_{m0}$ prior information on the inferred dark energy parameters $\omega_0$ and $\omega_a$. We first examine the interplay between observable sensitivities and measurement uncertainties. We then quantify how shifts in the prior mean of $\Omega_{m0}$ affect the posterior estimates and associated uncertainties of the DE parameters.

\subsection{Interplay of Sensitivities and Covariance Structure}
\label{subsec:sens_cov_results}

Among BAO combinations, $D_H/r_d$, $D_M/r_d$, and $D_V/r_d$ are most sensitive to variations in $\Omega_{m0}$, $\oo$, and $\oa$. Their dependence partly enters through $R(z)$ (the baryon-to-photon ratio), so analyses that include BBN priors on $\Omega_b h^2$ can indirectly modulate these sensitivities and the resulting constraints. We now comment on this linkage to strengthen the physical interpretation of Fig.~1.

As discussed in Section~\ref{subsec:sensitivities}, the constraining power of BAO observables depends on both their sensitivity to cosmological parameters and the magnitude of their measurement uncertainties, encoded in the covariance matrix. Figure~\ref{fig:sensitivities} shows the absolute sensitivities $|\partial O / \partial \theta|$ of BAO observables with respect to $\Omega_{m0}$, $\omega_0$, and $\omega_a$, while Equation~\eqref{eq:cov_matrix_block} and the covariance matrices in Section~\ref{subsec:mock_data} capture their observational errors.

For optimal parameter constraints, an observable should be highly sensitive to the parameter of interest while also exhibiting low measurement variance. In Fisher analysis, larger (squared) sensitivities and lower covariance elements both enhance the constraining power.

For example, QSO measurements at $z = 1.484$ show large covariance elements (high uncertainty), yet the $D_V/r_d$ observable exhibits strong sensitivity to $\Omega_{m0}$ at this redshift (Figure~\ref{fig:sensitivities}). Despite its noise, this observable contributes useful information on $\Omega_{m0}$. Conversely, $D_M/D_H$ shows generally low sensitivity to all parameters, and thus contributes little to their constraints even when its variance is moderate.

Notably, $D_H/r_d$ shows a non-monotonic sensitivity to $\Omega_{m0}$: high at low $z$, dropping around $z \sim 1.5$, and increasing again at high $z$. In contrast, its sensitivity to $\omega_0$ and $\omega_a$ increases more uniformly with redshift. This suggests that the most informative observable varies across the redshift range, depending on which parameter is being constrained.

\subsection{Bias in Parameter Estimates}
\label{subsec:bias_estimates}

In flat $\Lambda$CDM, we acknowledge a mild ($\sim1.8\sigma$) discrepancy between DESI DR2 BAO ($\Omega_{m0}=0.2975\pm0.0086$) and Pantheon+\,\&\,SH0ES ($\Omega_{m0}=0.334\pm0.018$). When switching to flat $\oo \oa$CDM, this tension in $\Omega_{m0}$ largely subsides, whereas a mild tension appears in $\oo$ instead. We now present both model cases side by side and emphasize that the prior-bias mechanism studied here applies precisely because of the parameter degeneracies connecting $\Omega_{m0}$, $H_0$, and the dark-energy parameters $(\oo, \oa)$.

We now present the key result of our study: the bias in DE parameter estimates introduced by an incorrect prior mean on $\Omega_{m0}$. Using 1000 mock DESI DR2 BAO datasets generated under a fiducial $\Lambda$CDM cosmology ($\Omega_{m0} = 0.3$, $\omega_0 = -1.0$, $\omega_a = 0.0$), we vary the mean of a Gaussian prior on $\Omega_{m0}$ from 0.27 to 0.33 and examine the resulting posterior estimates of $\omega_0$ and $\omega_a$.

\begin{figure*}
  \centering
  \begin{subfigure}[b]{0.48\textwidth}
    \includegraphics[width=\textwidth]{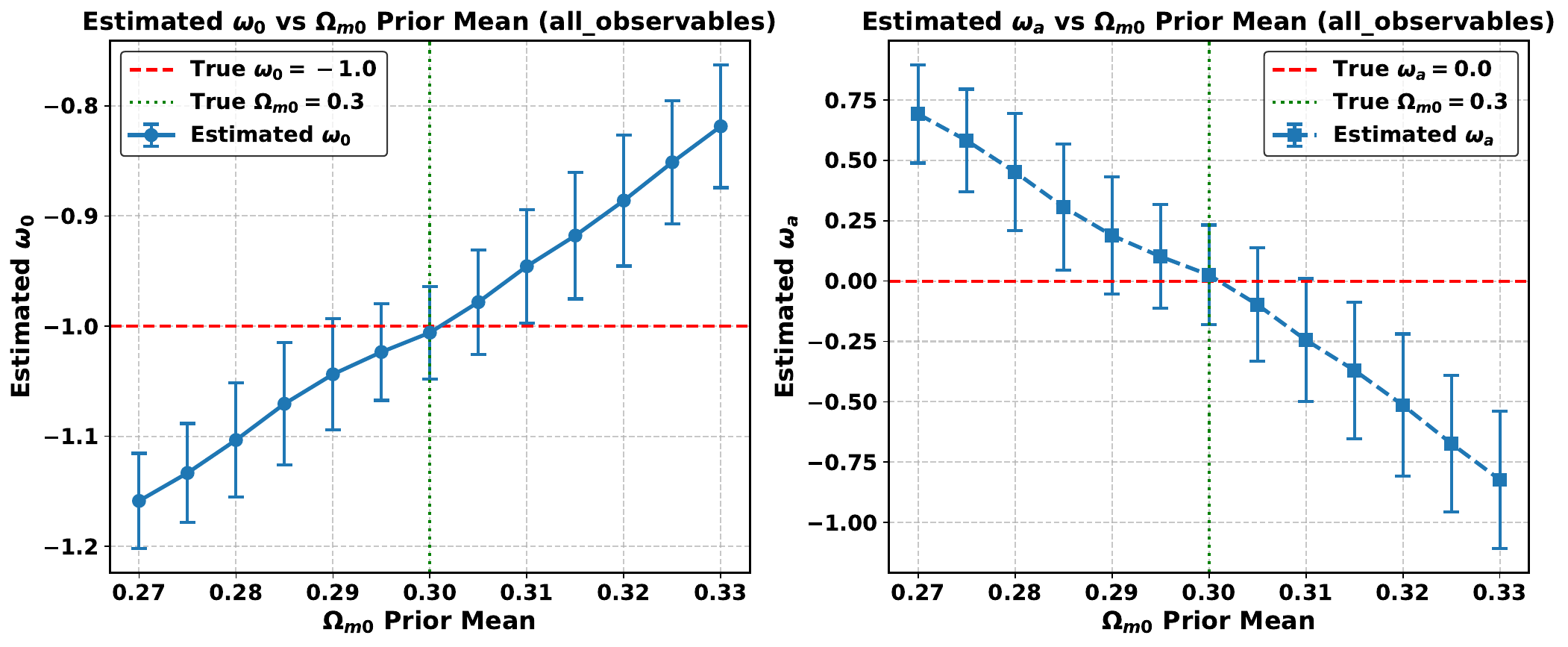}
    \caption{All observables}
    \label{fig:est_all}
  \end{subfigure}
  \begin{subfigure}[b]{0.48\textwidth}
    \includegraphics[width=\textwidth]{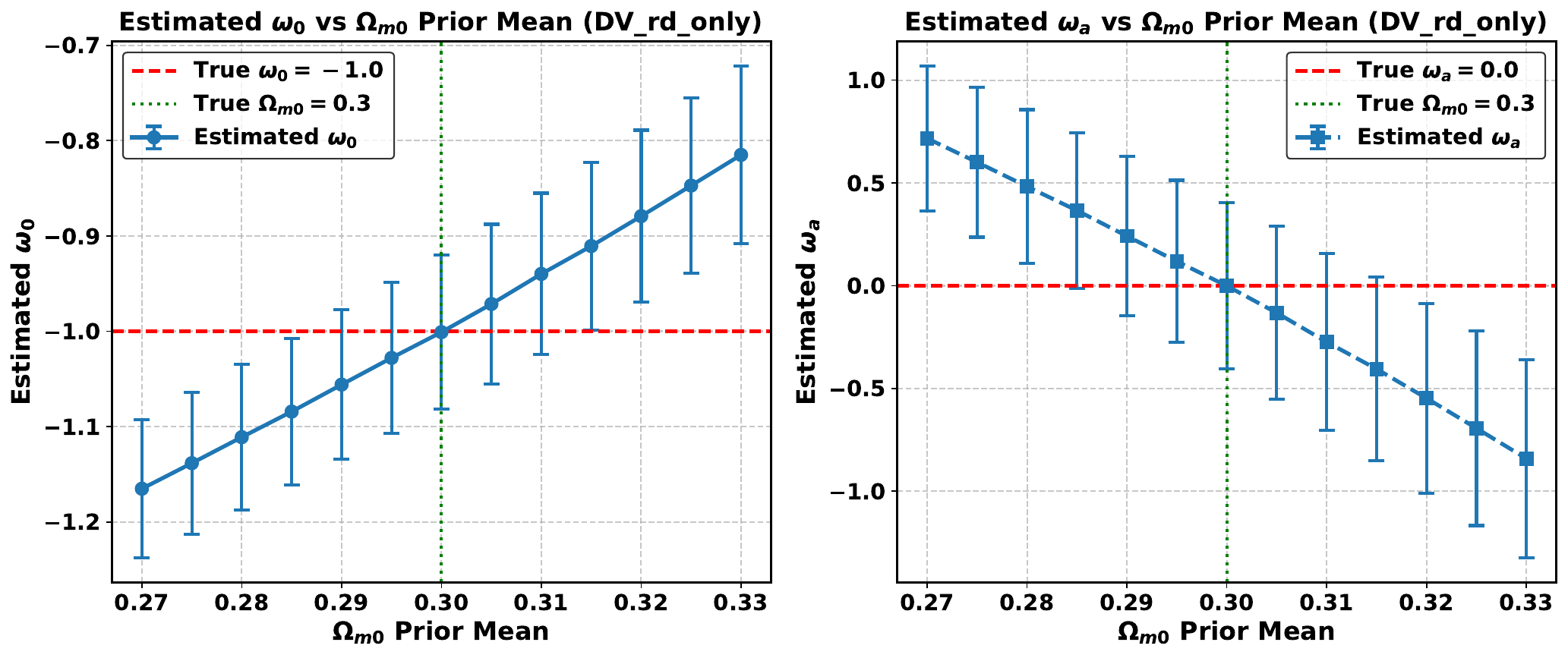}
    \caption{$D_V/r_d$ only}
    \label{fig:est_DVrd}
  \end{subfigure}
  \begin{subfigure}[b]{0.48\textwidth}
    \includegraphics[width=\textwidth]{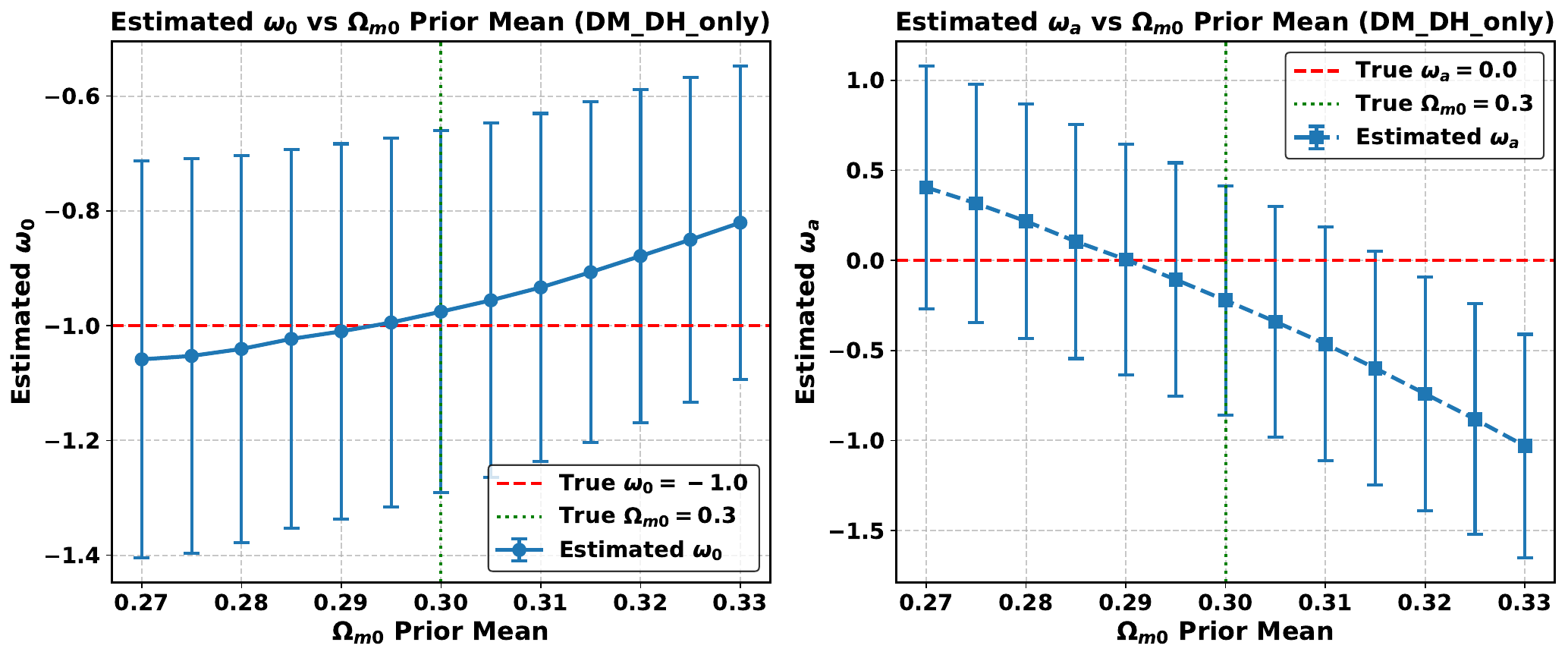}
    \caption{$D_M/D_H$ only}
    \label{fig:est_DMDH}
  \end{subfigure}
  \begin{subfigure}[b]{0.48\textwidth}
    \includegraphics[width=\textwidth]{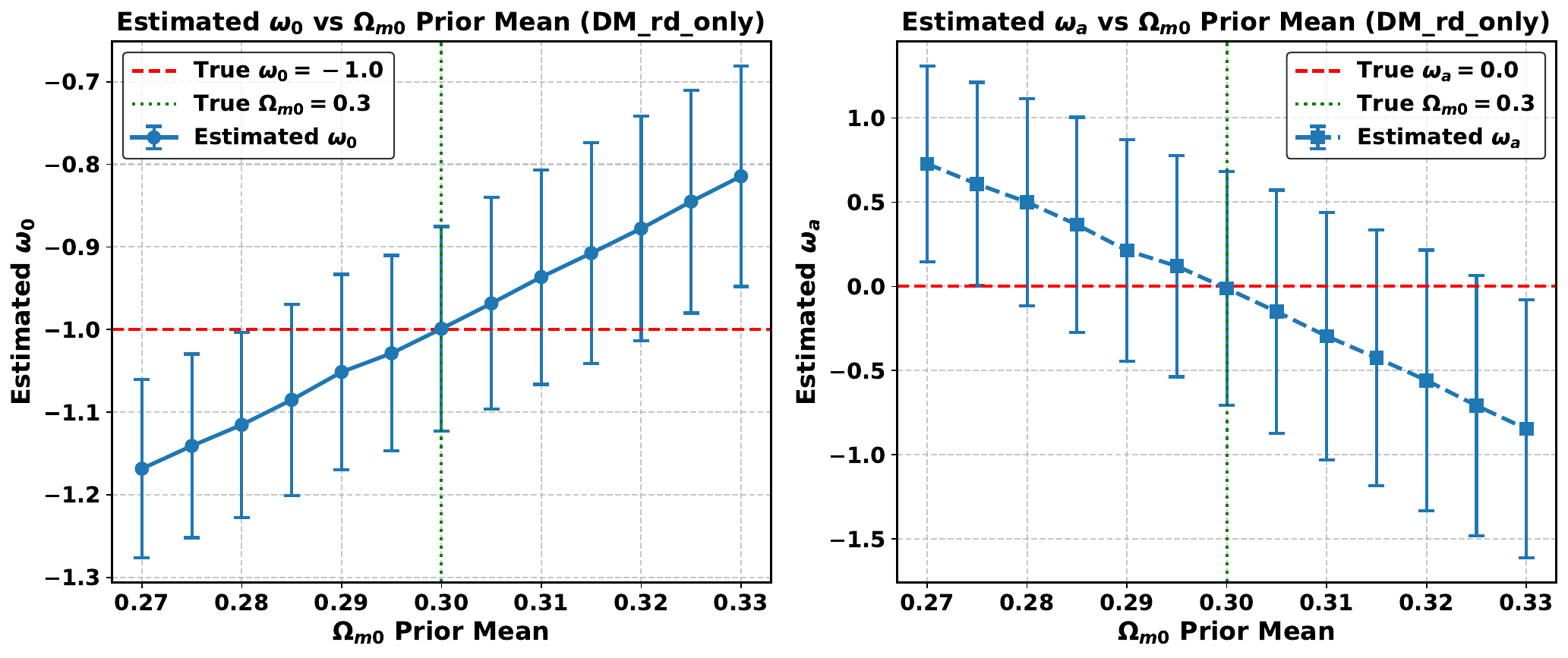}
    \caption{$D_M/r_d$ only}
    \label{fig:est_DMrd}
  \end{subfigure}
  \begin{subfigure}[b]{0.48\textwidth}
    \includegraphics[width=\textwidth]{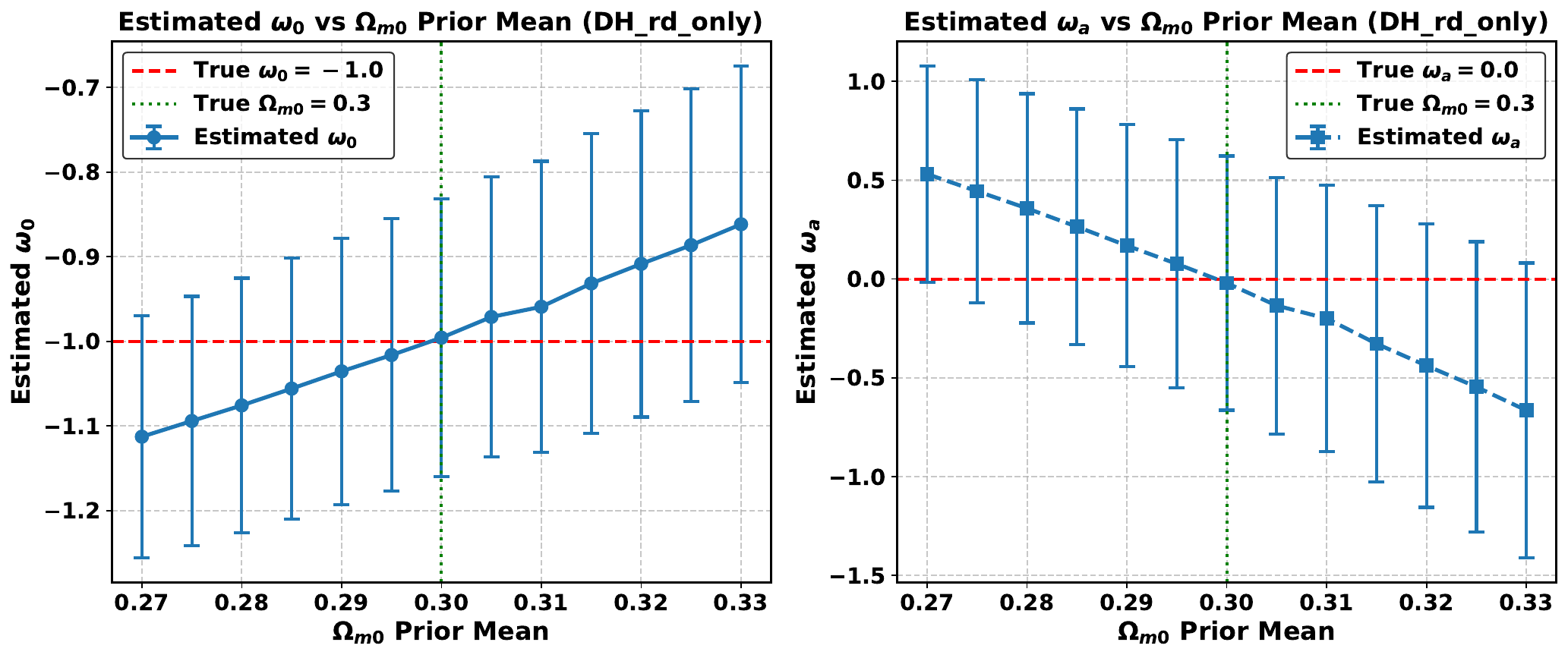}
    \caption{$D_H/r_d$ only}
    \label{fig:est_DHrd}
  \end{subfigure}
  \caption{Estimated $\omega_0$ and $\omega_a$ as a function of the $\Omega_{m0}$ prior mean for different combinations of DESI DR2 BAO observables.  Points show the mean of best-fit (or posterior-mean) values across 1000 mock realizations; error bars denote the standard deviation across realizations (not single-mock posterior widths).  Red dashed lines show the true $\Lambda$CDM values, and the vertical green dotted line indicates the true $\Omega_{m0} = 0.3$.}
  \label{fig:estimated_params_comparison}
\end{figure*}

Figure~\ref{fig:estimated_params_comparison}(a) shows a clear linear bias trend: as the $\Omega_{m0}$ prior mean deviates from the true value (vertical green line), the estimated $\omega_0$ and $\omega_a$ deviate from their fiducial values (horizontal red lines). This bias arises from degeneracies among the parameters, as the MCMC analysis finds a best-fit compromise between the data likelihood and the prior. Even though the mock data are generated from $\Lambda$CDM, an inconsistent prior leads to significantly non-$\Lambda$CDM posterior estimates.

For instance, applying a prior with mean $\Omega_{m0} = 0.33$ yields biased values $\omega_0 \approx -0.82 \pm 0.06$ and $\omega_a \approx -0.82 \pm 0.4$, which qualitatively resemble the results of combined analyses such as DESI+Pantheon+ ($\omega_0 = -0.888^{+0.055}_{-0.064}$, $\omega_a = -0.17 \pm 0.46$). Meanwhile, applying a prior close to the true value (e.g., 0.30) recovers unbiased estimates.

Subfigures~\ref{fig:est_DVrd} through \ref{fig:est_DHrd} highlight the role of individual observables. $D_V/r_d$ and $D_M/r_d$ are most influential in producing the bias trend due to their strong sensitivity to $\Omega_{m0}$. Conversely, $D_M/D_H$ contributes little bias, consistent with its low sensitivity in Figure~\ref{fig:sensitivities}.

\subsection{Uncertainty Propagation}
\label{subsec:error_propagation}

\begin{figure*}
  \centering
  \begin{subfigure}[b]{0.48\textwidth}
    \includegraphics[width=\textwidth]{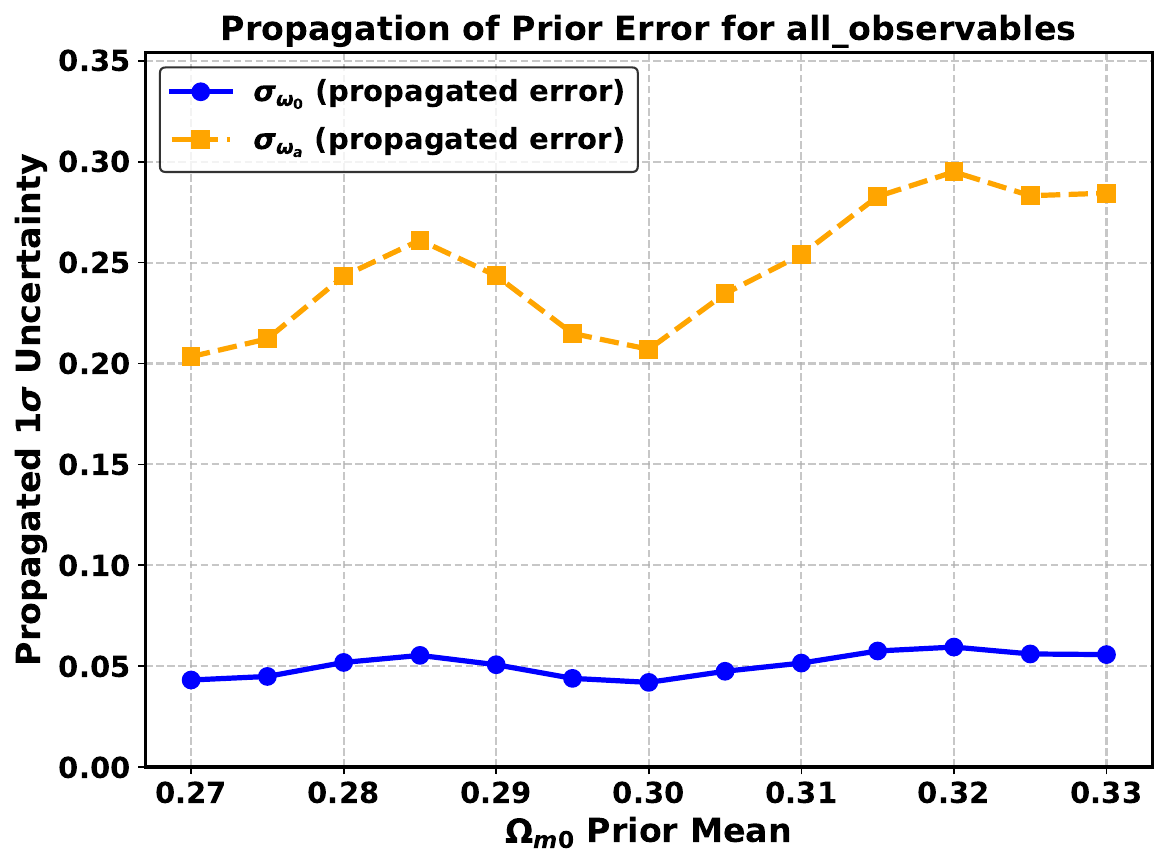}
    \caption{All observables}
    \label{fig:ep_all}
  \end{subfigure}
  \begin{subfigure}[b]{0.48\textwidth}
    \includegraphics[width=\textwidth]{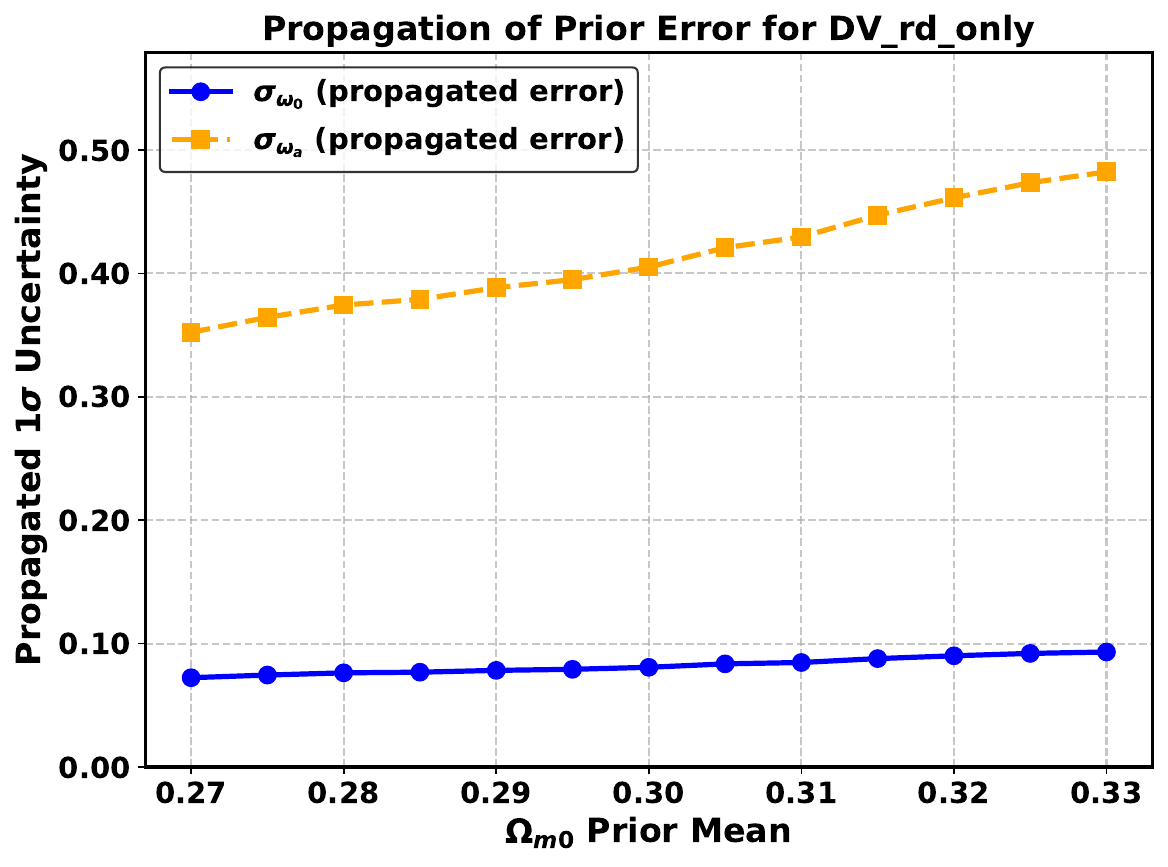}
    \caption{$D_V/r_d$ only}
    \label{fig:ep_DVrd}
  \end{subfigure}
  \begin{subfigure}[b]{0.48\textwidth}
    \includegraphics[width=\textwidth]{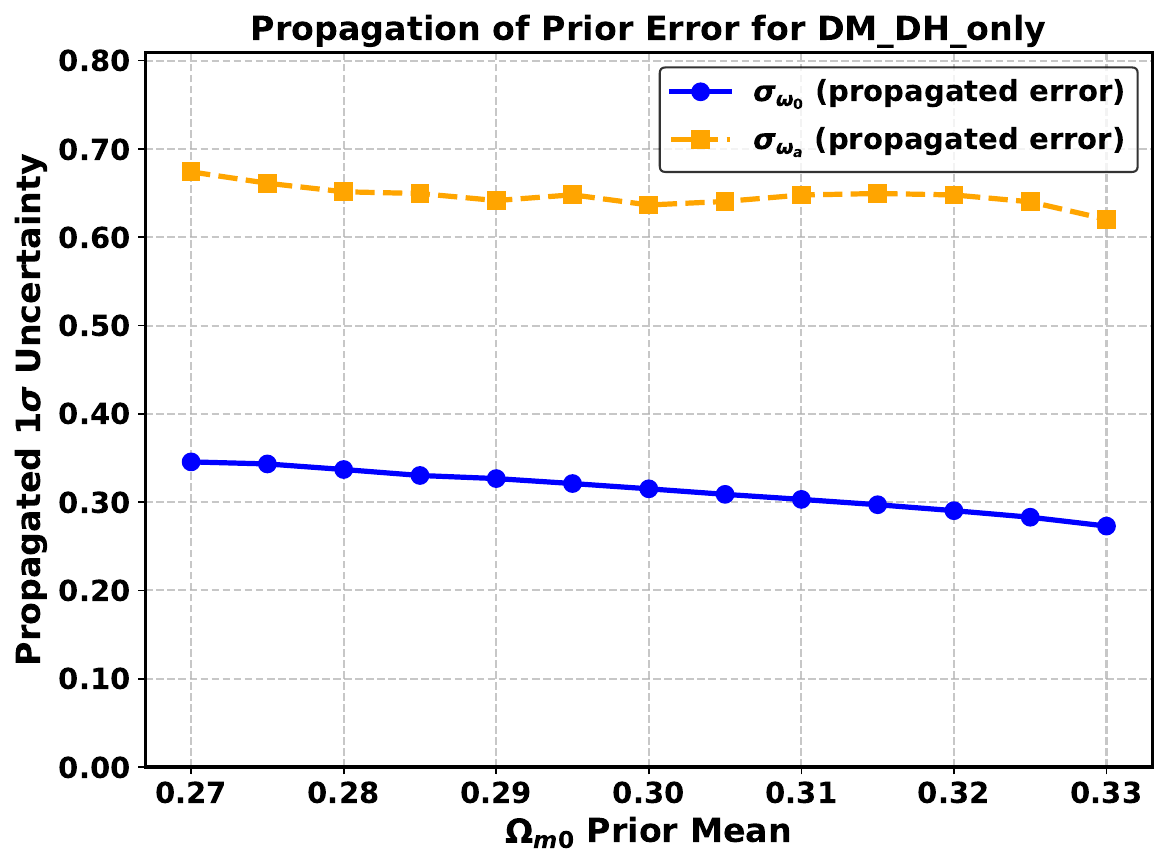}
    \caption{$D_M/D_H$ only}
    \label{fig:ep_DMDH}
  \end{subfigure}
  \begin{subfigure}[b]{0.48\textwidth}
    \includegraphics[width=\textwidth]{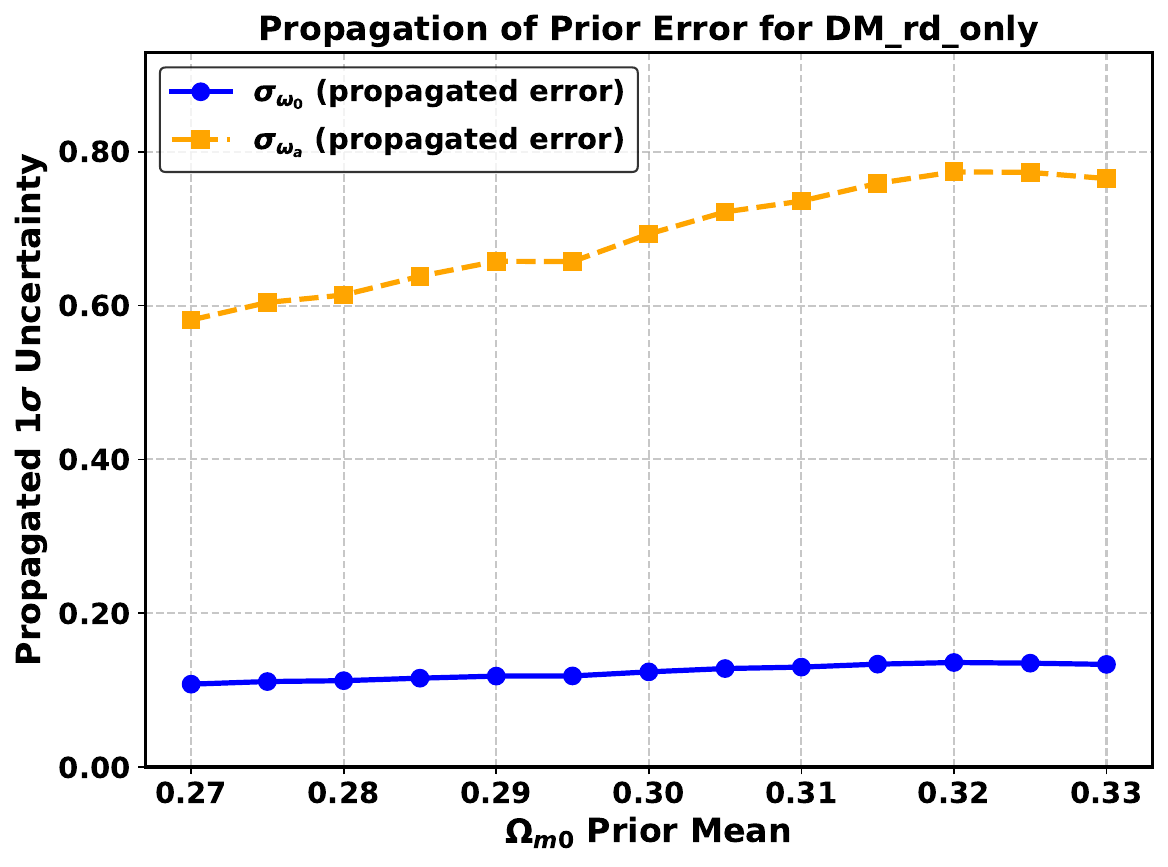}
    \caption{$D_M/r_d$ only}
    \label{fig:ep_DMrd}
  \end{subfigure}
  \begin{subfigure}[b]{0.48\textwidth}
    \includegraphics[width=\textwidth]{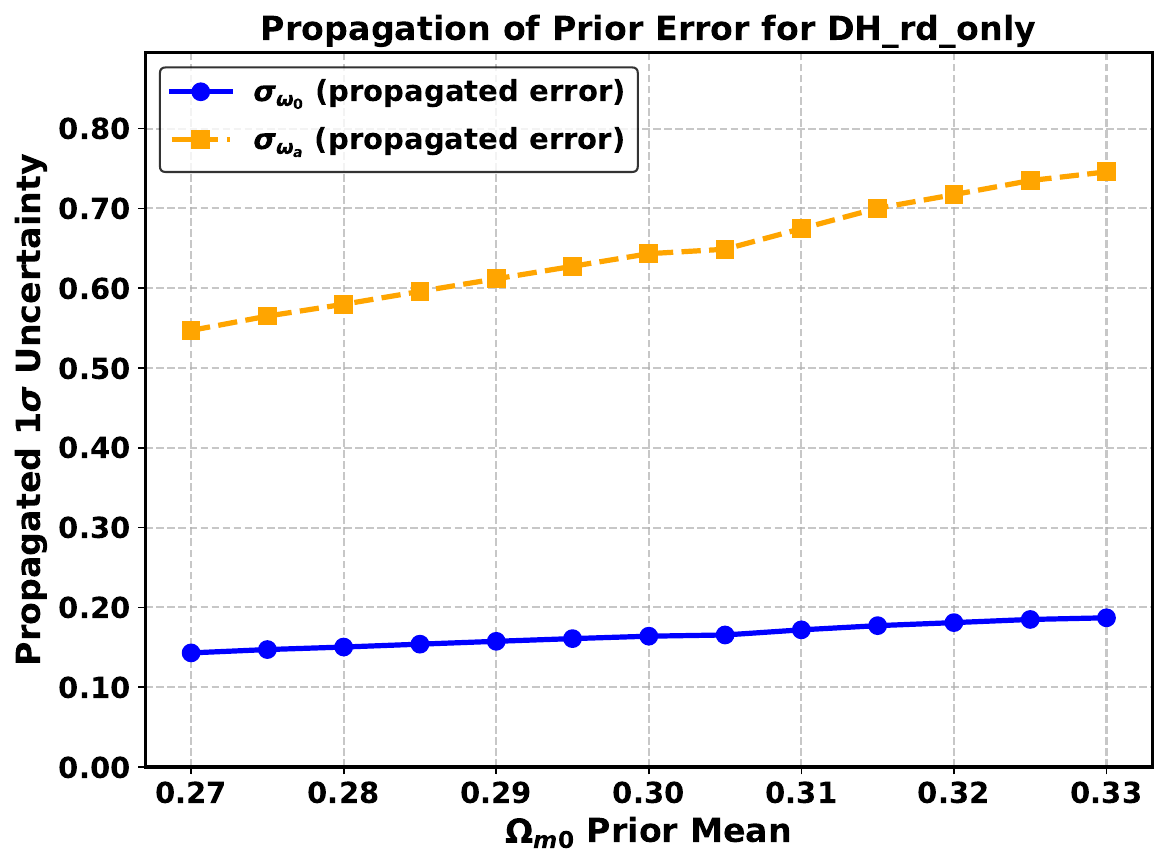}
    \caption{$D_H/r_d$ only}
    \label{fig:ep_DHrd}
  \end{subfigure}
  \caption{Propagated $1\sigma$ uncertainties on $\omega_0$ and $\omega_a$ as a function of $\Omega_{m0}$ prior mean. Each panel shows standard deviations from 1000 mock realizations for different BAO observables.}
  \label{fig:error_propagation_comparison}
\end{figure*}

We now assess how the $1\sigma$ uncertainties on $\omega_0$ and $\omega_a$ propagate as the $\Omega_{m0}$ prior mean changes. Figure~\ref{fig:error_propagation_comparison} displays the standard deviations of the estimated parameters across mock realizations.

A key observation is that the uncertainties are remarkably stable across the range of prior means (Figure~\ref{fig:ep_all}). This indicates that the precision of the DE parameters is primarily determined by the strength (i.e., width) of the prior, not its central value. Since we fix the prior variance in our simulations, the uncertainty on $\omega_0$ and $\omega_a$ remains largely unaffected by bias in the prior mean.

However, the inclusion of the $\Omega_{m0}$ prior significantly reduces the uncertainties compared to a BAO-only scenario. This occurs because the prior helps break degeneracies between $\Omega_{m0}$ and the DE parameters. Even if the prior is biased, it tightens the parameter ellipse by pinning down one direction in parameter space.

Individual observables differ in how strongly they benefit from the prior. For example, $D_V/r_d$-based constraints are sharper than those from $D_M/D_H$ (Figure~\ref{fig:ep_DMDH}), reflecting their relative sensitivities.

\subsection{Remark on mock design and observable redundancy}
\label{subsec:mock}
It is important to emphasize that the mock datasets employed in this study are not intended to provide an exact replication of the official DESI~DR2 likelihood.  Rather, they are designed as an experimental model that incorporates all major BAO observables $(D_V/r_d,\, D_M/D_H,\, D_M/r_d,\, D_H/r_d)$ to evaluate how the $\Omega_{m0}$ prior bias propagates to the DE parameters $(\omega_0,\,\omega_a)$.  In the actual DESI DR2 analysis, these observables are not fully independent: for instance, $D_V/r_d$ and $D_M/D_H$ are derived from the same radial and transverse measurements $(D_M/r_d,\,D_H/r_d)$,  so including all four simultaneously can lead to partial redundancy in the information content.  For this reason, a realistic MCMC inference based on DESI data should employ only one internally consistent basis,  such as $(D_M/r_d,\,D_H/r_d)$ or the isotropic $(D_V/r_d,\,D_M/D_H)$ representation,  depending on the redshift bin and tracer type.  The present mock analysis, by contrast, intentionally combines these observables to highlight general trends in parameter sensitivity and the direction of bias rather than to reproduce the exact DESI likelihood structure.

\section{Conclusion}
\label{sec:conclusion}

Because $\Omega_{m0}$ and $H_0$ are partially degenerate in simple background models, a biased prior on $\Omega_{m0}$ can propagate to effective shifts in $H_0$ and, through degeneracies, to $(\oo\,,\oa)$. This mechanism highlights the importance of prior-consistency checks in multi-probe combinations and provides additional context for the Hubble-tension discussion.

In this work, we have demonstrated that prior assumptions on the matter density parameter $\Omega_{m0}$ can introduce significant biases in the estimation of dark energy parameters $\omega_0$ and $\omega_a$, even when the true cosmology follows the $\Lambda$CDM model. Using 1000 mock DESI DR2 BAO datasets simulated under a fiducial $\Lambda$CDM cosmology, we systematically varied the mean of a Gaussian prior on $\Omega_{m0}$ and analyzed the resulting impact on posterior parameter estimates.

Our results show that a prior mean deviating from the true value (e.g., 0.33 instead of 0.30) leads to biased estimates of $\omega_0$ and $\omega_a$, shifting them away from the $\Lambda$CDM predictions. This effect arises from the degeneracies between $\Omega_{m0}$ and the DE equation-of-state parameters, and it persists even when the observational data themselves are unbiased. The induced bias causes the posteriors to resemble those reported in current combined analyses of DESI DR2 BAO and SNe data, where non-$\Lambda$CDM values of $\omega_0$ and $\omega_a$ have been observed.

Importantly, we also find that the inclusion of such a prior—while introducing bias—significantly reduces the $1\sigma$ uncertainties of the DE parameters. This can give a false impression of statistical robustness, as the biased estimates may appear well constrained. Thus, precision may be achieved at the expense of accuracy if prior tensions are not properly addressed.

Similar assessments of prior dependence have been discussed in alternative frameworks such as $f(T)$ or teleparallel gravity~\cite{Briffa:2021nxg}, highlighting that the issue of prior consistency is broadly relevant across extended cosmological models.

These findings suggest that the apparent deviations from $\Lambda$CDM in recent cosmological analyses may not require new physics, but could instead be artifacts of mismatched priors. Our study underscores the importance of carefully assessing the consistency of prior information in multi-probe analyses. Persistent tensions in $\Omega_{m0}$ measurements across datasets represent a substantial source of systematic uncertainty. Future work should prioritize identifying the origin of these discrepancies and developing methodologies to account for them in joint parameter inference frameworks, in order to obtain unbiased and reliable constraints on the nature of dark energy.



\section*{Data Availability}

The authors confirm that the data supporting the findings of this study are available within the references and their supplementary materials. This study uses only published summary statistics from DESI DR2 (Table IV) and simulated mock datasets derived from them. The scripts and configuration used to generate the figures are available from the corresponding author upon reasonable request

\bibliographystyle{mnras}
\bibliography{references}

\end{document}